\documentclass[11pt]{article}
\usepackage{amsmath,amsthm,latexsym,amssymb,amsfonts,epsfig}
\usepackage{fontenc,dsfont,layout}

\numberwithin{equation}{section}

\newcommand{\R}{\mathbb{R}}                  
\newcommand{\C}{\mathbb{C}}               
\def\H{{\cal H}}

\DeclareMathOperator{\porder}{\mathcal{P}}             

\newcommand{\bra}[1]{|{\bf{#1}}\rangle}                        
\newcommand{\ket}[1]{\langle{\bf{#1}}}
\newcommand{\one}{\text{\bf 1}}               

\def\baza{{}^{\textrm{\tiny{0}}} \! e}
\def\bazad{{}^{\textrm{\tiny{0}}} \! \omega}

\newcommand{\M}{\Sigma}
\def\q{{}^o\!q}
\def\e{{}^o\!e}

\def\SU(2){{\rm SU(2)}}
\newcommand{\be}{\begin{equation}}
\newcommand{\ee}{\end{equation}}

\newcounter{mnotecount}[section]

\begin{document}

\title{Closed FRW model in Loop Quantum Cosmology}
\author{{\L}ukasz Szulc$^{1}$\thanks{lszulc@fuw.edu.pl}, Wojciech
Kami\'nski$^{1}$\thanks{wkaminsk@fuw.edu.pl},
        Jerzy Lewandowski$^{1,2}$\thanks{lewand@fuw.edu.pl}}
\date{\it 1. Instytut Fizyki Teoretycznej,
Uniwersytet Warszawski, ul. Ho\.{z}a 69, 00-681 Warszawa, Poland\\
2. Physics Department, 104 Davey, Penn State, University Park, PA 1602, USA \\
[.5cm]} \maketitle
\begin{abstract}
The basic idea of the LQC applies to every spatially homogeneous
cosmological model, however only the spatially flat (so called
$k=0$) case has been understood in detail in the literature thus
far. In the closed (so called: k=1)  case certain technical
difficulties have been the obstacle that stopped the development.
In
this work the
 difficulties are overcome,  and a new LQC model of the spatially
 closed, homogeneous, isotropic universe is constructed. The
 topology of the spacelike section of the universe is
assumed to be that of SU(2) or SO(3). Surprisingly, according to the
new results achieved in this paper, the two cases can be
distinguished from each other just by the local properties of the
quantum geometry of the universe! The quantum hamiltonian operator
of the gravitational field takes the form of a difference operator,
where the elementary step is  the quantum of the 3-volume derived in
the flat case by Ashtekar, Pawlowski and Singh.  The mathematical
properties of the operator are studied: it is essentially
self-adjoint,  bounded from above by $0$, the $0$ itself is not an
eigenvalue, the eigenvectors form a basis. An estimate on the
dimension of the spectral projection on any finite interval is
provided.
\end{abstract}

%

\section{Introduction}
\quad \quad   Loop Quantum Cosmology is the Loop Quantum Gravity
\cite{1,2,3} motivated approach to the quantization of the symmetric
cosmological models. It was started by Martin Bojowald \cite{4,5,6}.
The simplification used in the  quantum cosmological models
framework consists in reducing the phase space of the full theory to
the finite dimensional phase space  of degrees of freedom of a given
cosmological model. That idea had been applied long before the LQC
\cite{7,8}. The new strategy introduced by Bojowald is to maintain
in the construction of a quantum cosmological model the structure of
the Loop Quantum Gravity \footnote{See review \cite{26}}. The
homogeneous, isotropic case is  best understood these days. Already
the first calculations \cite{9,10} provided qualitatively new
insight into the quantum structure of spacetime near the event
classically known as the Big-Bang singularity. The LQC evolution
does not break down when the scale factor vanishes. However, these
works were incomplete in that the physical sector of the theory had
not been completed. This situation was rectified in \cite{14,13}.
Furthermore, a more careful implementation of the physical
considerations of full LQG \cite{12,13,14} - in particular using the
minimal quantum of 2-area \cite{11} - finally led  to the improved
Hamiltonian constraint of Ashtekar, Pawlowski and Singh (APS)
\cite{15}. The key difference between this quantum dynamics  and the
one used previously is that each step in the resulting quantum
evolution increases the 3-volume of the spacelike slice by a fixed,
always the same, amount. This emergence of the elementary quantum of
volume is a purely dynamical effect - the eigenvalues of the
kinematical quantum volume operator take all the possible values
from $0$ to $\infty$.

The outlined result of APS concerns the spatially flat case of the
homogeneous, isotropic universe (so called k=0 case). Whereas the
basic ideas of the (improved as well as unimproved) LQC approach
easily apply to the closed ($k=1$) or hyperbolic ($k=-1$) cases
\cite{16,17}, some technical difficulties were actual obstacle in
completing the quantization task. In this work we overcome  the
difficulty and construct the LQC model of the spatially closed,
homogeneous, isotropic universe. The topology of the spacelike
section of the universe is assumed to be that of SU(2) or SO(3). The
significant  result is that the
 quantum hamiltonian operator of the gravitational field (that is the gravitational
 part of the scalar constraint) takes the form of
the difference operator, where the elementary step is again the
quantum of the 3-volume derived in the flat case by APS. Next we
study the properties of the quantum hamiltonian operator defined in
the kinematical Hilbert space of gravitational degrees of freedom.
We show this operator is bounded from above, is essentially
self-adjoint on its natural domain and its self-adjoint extension
has trivial kernel. The Hilbert space admits decomposition into
subspaces preserved by the operator, so called super selection
sectors. In each of the super selection sectors the operator has a
discrete spectrum.
 The difference between the SU(2) and SO(3) cases consists  in
 different values of the
 parameter used through out the work. In the last section we point out
 the surprising local difference between the quantum geometries of
 the two globally different cases.

In this work we are concerned with the gravitational field degrees
of freedom only.  The full model should be defined in the tensor
product of the Hilbert space we consider below and the Hilbert space
of a given matter field excitations. Then, the full hamiltonian is
the gravitational hamiltonian plus the matter term. The results
established in this paper still apply to the gravitational part of
the hamiltonian operator. They will be used in the next paper
\cite{18} to provide exact analytic approach to the full quantum
scalar constraint equation.

In the meantime, APS plus  Vandersloot  \cite{19} have also
constructed the LQC model for the spatially closed, homogeneous,
isotropic universe. The gravitational parts of our models are
equivalent (except that we are not sure whether or not \cite{19}
admits the SO(3) case, but that is not a big deal) including the
factor ordering in the quantum hamiltonian operator \footnote{We
thank Abhay Ashtekar for the hint}. Moreover, the APS model includes
the massless scalar field and their work contains the full quantum
solution.

\section{Kinematics}
\subsection{The Ashtekar connection variables}
We consider now the Ashtekar phase space for spatially homogeneous,
isotropic cosmologies. We will focus only on the case $k=1$, where
the symmetry group ${\cal S}$ has the same Lie algebra as the
isometry group of the round, $3$-dimensional sphere $S_3$. We assume, the underlying manifold $G$ admits the structure of a Lie
group, either SU$(2)$ or SO$(3)$. We fix on $G$  either of these Lie
group structures. The symmetry group ${\cal S}$ takes the form
${\cal S}=G\times G$, where the left/right factor acts on $G$ by the
left/right action.

The Ashtekar phase space $\Gamma^{\cal S}_{\rm grav}$ of the
gravitational degrees of freedom of the cosmologies considered in
this paper is a subspace of the space $\Gamma$ of pairs $(A,\, P)$
of fields on the 3-manifold $G$, where $A$ is an su$(2)$ valued
$1$-form and $P$ an su(2)${}^*$ valued vector field of density
weight 1.  A pair $(A^\prime,\, P^{\prime})$ on $\M$ is said to be
spatially homogeneous and isotropic or, for brevity,
\emph{symmetric} if for every $s\in {\cal S}$ and for every $x\in\M$
there exists a neighborhood ${\cal U}_x$ and a gauge transformation
$g:{\cal U}_x \rightarrow SU(2)$, such that
\be \label{ge} (s^\star A^\prime, \, s^\star P^\prime)\ =\
(g^{-1}A'g\, + \, g^{-1}dg, g^{-1}P'g), \ee
on ${\cal U}_x$. An example of a symmetric pair can be constructed
as follows. Let
\be {}^o\!A\ =\ \omega_{\rm MC}\ee
where $\omega_{\rm MC}$ (
the Maurer-Cartan form) is a Lie algebra isomorphism which maps the
Lie algebra of left invariant vector fields on $G$ into the Lie
algebra su$(2)$. Let
$$e_{\rm MC}\ =\ ((\omega_{\rm MC})^*)^{-1},$$
hence $e_{\rm MC}$ is a left invariant vector field on $G$ taking
values in su$(2)$. To turn it into a density, let us fix an
invariant scalar product $\eta$ in su$(2)$,
$$ \eta(\xi,\zeta)\ =\ -2{\rm Tr}(\xi\zeta).$$
It induces, in a natural way, a left and right invariant metric tensor $\q$
on $G$. We will use the density of the weight 1, namely $\sqrt{\det
\q}$, to define
\be{{}^o\!P}\ =\ \sqrt{\det\q}\,e_{\rm MC}.\ee
An important geometric fact is, that for each symmetric pair
$(A',\, P')$ there is a globally defined gauge transformation
$g:\M\rightarrow \SU(2)$, such that the gauge transformed pair
\be (A,P) \ =\ (g^{-1}A'g\, + \, g^{-1}dg, g^{-1}P'g) \ee
has the following simple form

\be\label{ss} A\ =\ {c}\,\,{}^o\!A,\quad\  {P}\ =\
{\underbar{p}}\,\,{}^o\!P,\ee
where ${c}$ and $\underbar{p}$ are constants carrying the only
non-trivial information contained in the pair $(A^\prime, E^\prime)$
(the under-bar will be removed after the suitable rescaling).

Define  ${\bf \Gamma}^S_{\rm grav}$ to be the set of pairs
(\ref{ss}). The variables $({c}, \underbar{p})$ form a globally
defined coordinate system. The symplectic form $\Omega^{\cal S}_{\rm
grav}$ on $\Gamma^{\cal S}_{\rm grav}$ is given by the pullback of
the symplectic form $\Omega_{\rm grav}$ \be \Omega_{\rm
grav}(\delta_1,\delta_2)\ =\ \int_{G}\delta_1A^i_a\wedge
\delta_2P^a_i - \delta_2A^i_a\wedge \delta_1P_i^a\ee
of the full theory. The result is
\be \Omega^{\cal S}_{\rm grav}\ =\ 3V_0\,d{c}\wedge d\underbar{p},
\ee
where
%
%
\begin{equation}
V_0=\int_{G}\sqrt{\det\q}\,d^3 x =\left\{
\begin{array}{lll}
16\pi^2 & {\rm for} & G={\rm SU}(2)\\
8\pi^2  & {\rm for} & G={\rm SO}(3)\\
\end{array}\right.\label{V_0}
\end{equation}
is the volume of the 3-manifold $G$ with respect to the  metric
 $\q$.
The vector density $P$ defines the physical metric tensor $q$ on
$G$. Let $\tau_1,\tau_2,\tau_3$ be a basis of su$(2)$ orthonormal
with respect to the scalar product $\eta$, and
$\tau^1,\tau^2,\tau^3$ the dual basis of su$(2)^*$. The orthonormal
frame $e_1,e_2,e_3$ tangent to $G$, corresponding to
$P=P_i\otimes\tau^i$ and $q$, is determined as follows
\be \sqrt{\det q}e_i\ = \kappa\gamma P_i, \ee
where
$$\kappa = 8\pi G,$$
and $G$ is the Newton constant.
 Therefore, the physical meaning of the variable
$\underbar{p}$ is
$$ V\ =\ V_0(\kappa\gamma|\underbar{p}|)^{\frac{3}{2}}, $$
where $V$ is the physical volume of $G$ defined by the physical
metric tensor. It leads to the following rescaling
\be p =(V_0)^{\frac{2}{3}}\kappa\gamma \underbar{p},\ee
upon which
$$ |p|\ =\ V^{\frac{2}{3}},$$
and the sign of $p$ defines the orientation. The  Poisson bracket
between two functions $f,g\in C(\Gamma^{\cal S}_{\rm grav})$ is:
\begin{equation}
\{f,g\} =
\frac{\kappa\gamma}{3\ell_0}(\partial_cf\partial_pg-\partial_pf\partial_cg),
\ \ \ \ell_0\ =\ (V_0)^{\frac{1}{3}}.\label{non_zero_poisson}
\end{equation}

\subsection{The loop quantization}
In LQG the connection variable is replaced by the parallel transport
variable. In the homogeneous  case we consider the parallel
transports along finite segments of geodesic curves in $G$, called
edges. Let $\alpha:[0,\mu]\rightarrow G$, $\mu\in \mathbb{R}$, be an
edge. There exists $\xi\in$su$(2)$, such that $\eta(\xi,\xi)\ =\ 1$, and
\be\label{edge} \alpha(t)\ =\ \alpha(0)e^{t\xi}.\ee
Along the edge $\alpha$ we consider the parallel transport defined
by the connection $A=c\omega_{\rm MC}$,
\begin{equation}\label{holonomy}
h_\xi^{(\mu)}\ =\ \porder \mathrm{exp}\left(- \int A \right)= e^{-\mu c \xi} = \cos
\frac{\mu c}{2}\one - 2 \sin \frac{\mu c}{2}\xi.
\end{equation}
The entries $h_\xi^{(\mu)}{}^M_N$ of the matrix $h_\xi^{(\mu)}$ are linear
combinations of the functions $e^{\pm i\mu\frac{c}{2}}$.


The Poisson bracket between the variable $p$ and the loop variable given by
(\ref{non_zero_poisson}) is:
\begin{equation}
\{ p , e^{i \frac{\mu c}{2}} \}= -i \frac{8\pi G \gamma}{6\ell_0} \mu
\cdot e^{i \frac{\mu c}{2}}
\end{equation}
where $\mu \in ]- \infty,\infty[ $. The quantum algebra of basic
operators is defined by the commutator:
\begin{equation}
[ \widehat{e^{i \frac{\mu c}{2}}},\hat{p} ]=- \frac{8 \pi
\mathrm{l}_{\mathrm{Pl}}^2 \gamma}{6\ell_0} \mu \widehat{e^{i
\frac{\mu c}{2}}}.
\end{equation}

For the kinematical Hilbert space we take  $\H^{kin} = L^2
(\mathbb{R}_{Bohr},d\mu_{Bohr})$, where $\mathbb{R}_{Bohr}$ denotes the
Bohr compactification of the real line. The space is the completion
of the vector space of the formal finite linear combinations of
elements of the basis $\{\bra{\mu}\ :\ \mu\in\R\}$, endowed with the
hermitian scalar product defined by
\begin{equation}
\ket{\mu_i} \bra{\mu_j} = \delta_{{\mu_i},{\mu_j}}.
\end{equation}
The action of the fundamental operators is defined as follows:
\begin{equation}
\hat{p} \bra{\mu} =\frac{8 \pi \mathrm{l}_{\mathrm{Pl}}^2
\gamma}{6\ell_0} \mu \bra{\mu} \quad \mathrm{and} \quad \widehat{e^{i
\frac{\tilde{\mu} c}{2}}}\bra{\mu} = \bra{\mu
+\tilde{\mu}}\label{catmu}
\end{equation}

\section{Dynamics}
\subsection{Classical picture}
The gravitational terms in the diffeomorphism and, respectively,
Gauss constraints read:
$$ C_{\rm G}^{\rm grav}(\Lambda)\ =\ \int_G\Lambda^i D_aP^a_i,\ \ \
C_{\rm diff}^{\rm grav}(\vec{N})\ =\ \int_G N^aF_{ab}^iP_i^b +
C_{\rm G}^{\rm grav}(\Lambda).
$$
where
\be  F\ =\ dA + A\wedge A,\ \ \ {\rm and}\ \ \ D_aP^a_i\ =\
\partial_aP^a_i - P^a_jA^k_a c^j{}_{ki} .\ee
Each pair $(A,P)$ defined in (\ref{ss}), satisfies automatically
\be\label{gauss-diff} D_aP^a_i\ =\ 0, \ \ \ {\rm and} \ \ \
F_{ab}^iP_i^b \ =\ 0. \ee
Hence, the latter property (\ref{gauss-diff}) follows from the symmetry
assumption only. The consequence is, that in the presence of matter,
the corresponding matter terms of the Gauss and, respectively,
diffeomorphism constraint have to vanish separately.

The gravitational part $C_{\rm sc}^{\rm grav}(N)$ of the scalar
constraint (the time evolution generator) has the following general
form
\begin{equation}\label{LQGscalar}C_{\rm sc}^{\rm grav}(N)\ =\ \frac{1}{2}\,
\sqrt{\frac{\gamma}{\kappa}}\,\int_G N
\frac{P^a_iP^b_j\epsilon^{ijk}}{\sqrt{|{\rm det} P|}}\left( F^k_{ab}
- (1+\gamma^2)\epsilon^k{}_{nm}K^n_aK^m_b\right)
\end{equation}
Substituting the variables (\ref{ss}) and $N={\rm const}$ one would
get the following simple formula
\begin{equation}
C_{\rm sc}^{\rm grav}(N)\ = -N \left(\frac{3\sqrt{|p|}}{8 \pi G
\gamma^2} \ell_0^2\left( - c + c^2 \right) + (1+\gamma^{-2})\frac{3
\ell_0^2 \sqrt{|p|}}{4 \cdot 8 \pi G }\right).
\end{equation}
%
%
\subsection{Quantum picture: preparation}
Our aim is to mimic the quantization scheme used in LQG. Therefore
we go back to the LQG form of the scalar constraint
(\ref{LQGscalar}). According to LQG, the curvature $F$ in the scalar
constraint should be expressed by the holonomy map, the parallel
transport functions along suitable closed loop. The functions are
well defined operators in the kinematical Hilbert space. Next, in
LQG,  each loop is shrunk to a point. The limit of the corresponding
scalar constraint operator exists in the dual space, the space of
diffeomorphism invariant linear functionals. In LQC, on the other
hand, the diffeomorphism constraint is solved on the classical
level,  the solutions are given in a  fixed gauge (we choose one
representative (\ref{ss}) from each diffeomorphism equivalence
class). As the consequence,  the quantum scalar constraint operator
has to be defined directly in the kinematical Hilbert space. To
preserve the analogy with LQG, in LQC we quantize the scalar
constraint {\bf } replacing the curvature components by suitably
chosen parallel transport operators.{\bf } However we stop shrinking
the loops at the stage when they reach the minimal, non-zero area
allowed by the LQG.

In what follows we use the following decomposition
\be \omega_{\rm MC}\ =\ \bazad^i \tau_i,\ \ \ \ \ \ \e_{\rm MC}\ =\
\baza_i\tau^{*i}. \ee
 Specifically, to a
component $F_{ab}\baza^a_i\baza^b_j$ we assign a two dimensional
plane
$$S_{ij}^{(\mu)}\ =\
\{e^{s\tau_i}e^{t\tau_j}: 0\le s,t\le \mu\},$$
and the parallel transport $h^{(\mu)}_{ij}$ along its contour
defined by the connection (\ref{ss}). The contour is a square,
consisting of the four geodesic edges of equal length, each
orthogonal to the neighbors. The parallel transports along the
segments can be calculated using
(\ref{holonomy}).\footnote{Seemingly, one of the edges $\alpha_3(t)\
=\ e^{-t\tau_i}\,e^{\mu\tau_i}e^{\mu\tau_j}$ is an integral curve of
a right invariant vector field, whereas (\ref{edge}) defines an
integral line of a left invariant vector field, however we should
remember that $e^{t\xi} g = g e^{t\xi'}$ where $\xi'= g^{-1}\xi g$.}
The result is
\begin{equation}
h_{ij}^{(\mu)}\ =\ e^{-(1-c)\mu \tau_j} e^{\mu c \tau_i} e^{(1-c)\mu
\tau_j} e^{-\mu c \tau_i}
\end{equation}
On the other hand, the curvature is
$$F^k_{ab}\baza^a_i\baza^b_j\ =  c^k{}_{ij}(- c + c^2),$$ and, as expected,
\begin{equation}
\lim_{\mu \to 0} \frac{2}{\mu^2} {\rm Tr}(h_{ij}^{(\mu)} \tau_k)=
F_{kab}\baza^i_a\baza^j_b.
\end{equation}
However, as we have already mentioned, we are not going to shrink
the loop to a point. Following \cite{15} we will stop the shrinking
at $\mu=\bar{\mu}$ when the physical area of the loop reaches the
minimal non-zero area eigenvalue $a_0$ \cite{11}, and replace the
curvature component by
\be \frac{2}{\bar{\mu}^2}{\rm Tr} h_{ij}^{(\bar{\mu})}\tau_k\ =\
\frac{1}{\bar{\mu}^2}\sin((c-1)\bar{\mu})\,\sin (c\bar{\mu})\,
c_{kij}. \ee

To calculate the physical area of the surface $S_{ij}^{(\mu)}$
note that its intrinsic geometry is  flat. Indeed, consider the
coordinate system $(s,t)$ defined on $S_{ij}^{(\mu)}$ and the
corresponding vector fields $\partial_s$ and $\partial_t$ tangent to
$S_{ij}^{(\mu)}$. They are commuting Killing vectors of the physical
metric $q$ defined on $G$. Therefore the components of the induced
metric tensor are constant on the surface due to the vanishing of
the following Lie derivatives
$$ {\cal L}_{\partial_A}q(\partial_B,\partial_C)\ =\ 0,\ \ \ \ \ \
A,B,C = t,s.$$ In the point of $G$ corresponding to the identity of
the group structure, it is easy to compute
$$q(\partial_A,\partial_B)\ =\ \frac{p}{\ell_0^2}\delta_{AB}.$$
Therefore, the area of the surface $S_{ij}^{(\mu)}$ is just
$$ \mathrm{Ar}= {\mu}^2 \frac{|p|}{\ell_0^2}.$$
This implies the following condition on the value  $\bar{\mu}$ of
the parameter $\mu$ is
\begin{equation}
\frac{|p|}{\ell_0^2} \cdot \bar{\mu}^2\ =\ a_0\ =\ 2\sqrt{3}\pi \gamma
\mathrm{l}_{\mathrm{Pl}}^2,
\end{equation}
hence
\be\label{mubar}\bar{\mu}\ = \sqrt{\frac{a_0}{|p|}}\,\ell_0 \ee
Alternatively to the square $S^{(\mu)}_{ij}$, we could consider
${S'}_{ij}^{(\mu)}\ =\ \{e^{s\tau_j}e^{t\tau_i}: 0\le s,t\le \mu\}$,
but the resulting replacement for the curvature is not sensitive to
that ambiguity.

Going back to the scalar constraint (\ref{LQGscalar}), we will still
take advantage of the special form (\ref{ss}) of our variables
$(A,P)$, namely, in the symmetric case we have the following extra
identity
\be 2K^i_{[a}K^j_{b]}\ =\
\frac{1}{\gamma^2}\epsilon^{ij}{}_kF^k_{ab}\ +\
\frac{1}{2}\bazad^i_{[a}\bazad^j_{b]}, \ee
where the second term is a constant (independent of the dynamical
variables).

Taking all that into account, we derive the following LQC
modification of the gravitational term of the scalar constraint:
\begin{equation}\label{scalar} C_{\rm sc}^{\rm grav}(N)
=-N \frac{3\ell_0^2}{8 \pi G \gamma^2}\sqrt{|p|}\left(
 \frac{\sin^2\,(\bar{\mu}(\frac{1}{2}-c))}{\bar{\mu}^2} -
\frac{\sin^2\,(\frac{1}{2}\bar{\mu})}{\bar{\mu}^2}+
\frac{1}{4}(1+\gamma^{2})\right),
\end{equation}
with $\bar{\mu}$ being defined in (\ref{mubar}). Intentionally we
wrote the latter expression in the way manifestly negative definite,
and we will preserve that  property by suitable quantization of the
constraint.

We are done as far as expressing the curvature is concerned. In LQG
we make one more trick: the factor $\frac{P^a_iP^b_j}{\sqrt{|{\rm
det}\,P}|}$ in (\ref{LQGscalar}) which requires the special care due
to the denominator, is expressed by functions written in the form
$h^{-1}\{h,V\}$, where $h$ is a parallel transport function (of the
variable $A$) and $V$ is the 3-volume  function (of $P^a_i$). In our
case all that factor is proportional to the $\sqrt{|p|}$, hence it
is well defined. However, for the sake of analogy with LQG, for the
quantization we will write $\sqrt{|p|}$ in the following exact form:
\begin{equation}
{\rm sgn}(p)\sqrt{|p|}\ =\ \frac{4\ell_0}{3\kappa \gamma \bar{\mu}}
\sum_k \mathrm{Tr} \left( h_k^{(\bar{\mu}) -1} \{
h_k^{(\bar{\mu})},V \}\tau_k\right).
\end{equation}
Consequently,  $\bar{\mu}$ is the one defined in (\ref{mubar}).
%
%
\subsection{Quantization}
After that preparation we are in the position to define the quantum
scalar constraint. However, as one could see, yet a new type of
functions has emerged, namely the function
$e^{i\frac{\bar{\mu}c}{2}}$. The quantization is tricky because
$e^{i\frac{\bar{\mu}c}{2}}$ involves both variables $c$ and $p$. We begin with
the observation that
$$\widehat{\exp({i\frac{kc}{2}})}\ \bra{\mu}\ =\
\bra{\mu+k}.$$
Hence the operator $\widehat{\exp({i\frac{kc}{2}})}$ is the pullback
induced by the translation map
$$\exp(k\frac{d}{d\mu})\ \colon\ \R\ni\mu\mapsto\mu+k\in\R$$
generated by the vector field $k\frac{d}{d\mu}$. Now, by analogy,
one  defines the operator $\widehat{\exp(i\frac{\bar{\mu}c}{2})}$ to
be the pullback induced  by the  map $\R\rightarrow\R$ generated by
the vector field $\bar{\mu}(\mu)\frac{d}{d\mu}$, that is by
$\exp(\bar{\mu}\frac{d}{d\mu})$.  This operator can be expressed
again as a translation, but in the different parametrization of
$\R$, namely  $\nu:\R\rightarrow\R$, such that
$$\bar{\mu}\frac{d}{d\mu}\nu\ =\ 1,$$
for example by
\begin{equation}
\nu\ =\ K{\rm sgn}(\mu)|\mu|^{\frac{3}{2}},\ \ \ \ \ \ \  K\ =\frac{2\sqrt{2}}{\ell_0 \sqrt{\ell_0} 3 \sqrt{3 \sqrt{3}}}
\end{equation}
To define the action of the operator
$\widehat{\exp(i\frac{\bar{\mu}c}{2})}$ we just need to relabel the
basis $\bra{\mu}$ is the following way
\begin{equation}\label{nu} |\nu)\ :=\
\bra{\mu}\,\bigg|_{\mu={\rm sgn}(\nu)(\frac{|\nu|}{K})^{\frac{2}{3}}},
\end{equation}
and set
\begin{equation}\label{holmubar}
\widehat{e^{i\frac{\bar{\mu}c}{2}}} |\nu) = |\nu +1).
\end{equation}
Remarkably, the parameter $\nu$ and the operator (\ref{holmubar})
have clear interpretation in terms of the physical 3-volume operator
$$\hat{V}=| \hat{p}|^{3/2} $$
Using $(2.16)$ and $(3.15)$ we get formula for eigenvectors  and
eigenvalues of the volume operator:
\begin{equation}\label{nu1}
\hat{V} |\nu) = V_1\,|\nu|\ |\nu),\ \ \ \ V_1 = \left( \frac{8 \pi
\gamma}{6 \ell_0} \right)^{\frac{3}{2}} l_{\mathrm{Pl}}^3 \frac{1}{K}.
\end{equation}
Hence, the operator
$\widehat{\exp(i\frac{\bar{\mu}c}{2})}$ is just the shift in the
volume eigenvalue for the unit $V_1$, increasing or decreasing the
volume  depending on the orientation of the frame $P$ (the sign of
$\nu$).

We also extend our definition in the obvious way:
\begin{equation}\label{U}
U_b |\nu) \ :=\ \widehat{e^{i b \frac{\bar{\mu}c}{2}}} | \nu) = |\nu +
b)
\end{equation}
where $b$ is an arbitrary real number.

We are now in the position  to define the operator corresponding to
the factor $\sqrt{|p|}$. Using $(3.13)$ we get:
\begin{equation}
 \widehat{ \sqrt{|p|} } |\nu) = \mathrm{sgn}(\nu)\frac{2 \ell_0}{8\pi G
\gamma \hbar \bar{\mu} } \left( \widehat{e^{-i
\frac{\bar{\mu}c}{2}}} \hat{V} \widehat{e^{i \frac{\bar{\mu}c}{2}}}
- \widehat{e^{i \frac{\bar{\mu}c}{2}}} \hat{V} \widehat{e^{-i
\frac{\bar{\mu}c}{2}}} \right) |\nu).
\end{equation}
This operator turns out to be diagonal in the basis $|\nu)$.
Equations $(3.18)$ and $(3.19)$ give us following formula:
\begin{equation}\label{sqrtp}
\widehat{ \sqrt{|p|} } |\nu) = A(\nu)\,|\nu),\ \ \ \ A(\nu)\ =\
\frac{3\ell_0}{8 \pi \gamma}  l_{\mathrm{Pl}} \left( \frac{8 \pi
\gamma}{6\ell_0} \right)^{\frac{3}{2}} \frac{|\nu|^{1/3}}{K^{1/3}}
\left| |\nu + 1 | - |\nu - 1 | \right| .
\end{equation}

The next component of the future quantum scalar constraint operator
corresponding to (\ref{scalar}) is an operator corresponding to the
factors $\sin\,(\bar{\mu}(\frac{1}{2}-c))\,/\,{\bar{\mu}}$. With the
due care, using the $U$ operator (see \ref{U}) we
define:
\begin{align}\label{sin}
\widehat{\frac{\sin\,(\bar{\mu}(\frac{1}{2}-c))}{\bar{\mu}}}\ &:=\
\frac{1}{2i}\left(U_{-2}
\widehat{\frac{\exp(\frac{i}{2}\bar{\mu})}{\bar{\mu}}}\ - U_2
\widehat{\frac{\exp(-\frac{i}{2}\bar{\mu})}{\bar{\mu}}} \right)\nonumber\\
\widehat{\frac{\exp(-\frac{i}{2}\bar{\mu})}{\bar{\mu}}} \ |\nu)\
&:=\
\frac{\exp(-\frac{i}{2}\bar{\mu}(\mu(\nu)))}{\bar{\mu}(\mu(\nu))}\,
|\nu).
\end{align}
Since this operator does not satisfy the reality conditions, it is
not symmetric. However it can be used to define the symmetric scalar
constraint operator, as follows (we take $N$ for the constant lapse
function equal 1):
\begin{equation}\hat{C}_{\rm sc}^{\rm grav} \ =\ -
\frac{3\ell_0^2}{8 \pi G \gamma^2}\left(
\widehat{\frac{\sin\,(\bar{\mu}(\frac{1}{2}-c))}{\bar{\mu}}}\,\widehat{\sqrt{|p|}}
\,\widehat{\frac{\sin\,(\bar{\mu}(\frac{1}{2}-c))}{\bar{\mu}}}^\dagger-
\widehat{\sqrt{|p|}}\widehat{\frac{\sin^2\,(\frac{1}{2}\bar{\mu})}
{\bar{\mu}^2}}+
\frac{1}{4}(1+\gamma^{2})\widehat{\sqrt{|p|}}\right),\label{Chat}
\end{equation}

\section{Properties of the quantum scalar constraint operator}

\noindent{\bf The (initial) domain.} The quantum scalar constraint
operator $\hat{C}_{\rm sc}^{\rm grav}$ has been defined by (\ref{U},
\ref{sqrtp},\ref{sin},\ref{nu},\ref{Chat}) in the domain
\begin{equation}\label{domain}
{\cal D}\ =\ \{ \Psi\in{\cal H}^{\rm kin}\ :\ \Psi=\sum_{i=1}^n
a_i\,|\nu_i), \ a_i\in\C,\ \nu_i\in\R, \ n\in\mathbb{N},  \},
\end{equation}
where  the  elements of the basis $\{\bra{\mu}\ :\ \mu\in\R\}$
(\ref{catmu}) were relabeled using the parameter
$\nu\colon \R\rightarrow\R$ (\ref{nu}), proportional to the eigenvalues of
the volume operator (\ref{nu1}).
\bigskip

\noindent{\bf The action.} The operator can be written in the form
of the difference operator,
\begin{equation}\label{C2}
\hat{C}_{\rm sc}^{\rm grav} \ =\ -\frac{3\ell_0^2}{32\pi G\gamma^2}(
C_0 + U_4C_4 + U_{-4}C_{-4}),
\end{equation}
where $C_{-4}, C_0, C_4$ are some functions of the variable $\nu$
acting by multiplication, that is  $|\nu)\mapsto C_I(\nu)|\nu)$,
$I=-4,0,4$ and $U_{\pm 4}$ are the shift operators defined in
(\ref{U}). Specifically,
\begin{align}
C_4(\nu)\ &=\
-\frac{e^{-i\bar{\mu}(\mu(\nu+2))}}{\bar{\mu}^2(\mu(\nu+2))}A(\nu+2)
\ \ \ \ \ {C_{-4}}(\nu)\ =\
-\frac{e^{i\bar{\mu}(\mu(\nu-2))}}{\bar{\mu}^2(\mu(\nu-2))}A(\nu-2)
\nonumber\\
C_0(\nu)\ &= \  \frac{A(\nu-2)}{\bar{\mu}^2(\mu(\nu-2))} +
\frac{A(\nu+2)}{\bar{\mu}^2(\mu(\nu+2))} + (1 - 4
\frac{\sin^2(\frac{\bar{\mu}(\mu(\nu))}{2})}{\bar{\mu}^2(\mu(\nu))}+\gamma^2)A(\nu)
\end{align}
Similar properties of the action are well known from the $k=0$ case.
 Due to them
the evolution equation involving the gravitational part of the
scalar constraint operator  turns into a difference equation, with
the step being a multiple of the volume difference $4V_1$
(\ref{nu1}) \cite{16}. It is an important result,  that those
features can be generalized to the $k=1$ case.
\bigskip

\noindent{\bf The upper bounds.} The operator $\hat{C}_{\rm sc}^{\rm
grav}$ is manifestly negative definite,
$$\hat{C}_{\rm sc}^{\rm grav} \le 0.$$
Even stronger inequalities hold in the domain ${\cal D}$, namely
(see (\ref{sqrtp})):
\begin{equation}
\hat{C}_{\rm sc}^{\rm grav} \ \le\ - \frac{3\ell_0^2}{32 \pi\
G}\widehat{\sqrt{|p|}},\label{upper}
\end{equation}
Indeed, the inequality follows from the obvious inequalities
\begin{align}
\widehat{\frac{\sin\,(\bar{\mu}(\frac{1}{2}-c))}{\bar{\mu}}}\,
\widehat{\sqrt{|p|}}
\,\widehat{\frac{\sin\,(\bar{\mu}(\frac{1}{2}-c))}{\bar{\mu}}}^\dagger
\ &\ge 0,\nonumber\\
 - \frac{\sin^2\,(\frac{1}{2}\bar{\mu}(\mu(\nu)))}
 {{\bar{\mu}^2}(\mu(\nu))} +
\frac{1}{4} \ge\ 0.
\end{align}
%
%

\noindent{\bf The self-adjointness.} The operator $\hat{C}_{\rm
sc}^{\rm grav}$ has been defined as manifestly symmetric operator.
Moreover:
\medskip

 \noindent{\bf Proposition 1 }{\it The operator $\hat{C}_{\rm sc}^{\rm grav}$
 defined in the domain ${\cal D}$ is essentially self adjoint.}
\medskip

\noindent We skip the proof to the end of this section. We denote by
${\cal D}(\hat{C}_{\rm sc}^{\rm grav})$ the self-adjoint extension
of the domain ${\cal D}$.

%
\bigskip
\noindent{\bf
Sharp negative definiteness of $\hat{C}_{\rm sc}^{\rm grav}$. }The
inequality (\ref{upper}) will be used in the spectral
 analysis of the operator $\hat{C}_{\rm sc}^{\rm grav}$.
\medskip

\noindent{\bf Proposition 2 }{\it The equation
$$\hat{C}_{\rm sc}^{\rm grav} \Psi\ =\ 0$$ has no nontrivialsolution for
$\Psi\in {\cal D}(\hat{C}_{\rm sc}^{\rm grav})$. Therefore the
scalar constraint is sharply negative operator on its extended
domain ${\cal D}(\hat{C}_{\rm sc}^{\rm grav})$,}
\[
\hat{C}_{\rm sc}^{\rm grav} \ <\ 0.\label{sharp}
\]
\bigskip

\noindent{\bf The decomposition into preserved subspaces}. We will
continue the analysis of the scalar constraint operator using
decomposition of the Hilbert space into separable, preserved
subspaces. The operator $\hat{C}_{\rm sc}^{\rm grav}$ preserves
every subspace
\begin{equation} {\cal
H}_{\epsilon}\ =\ {\rm Span}\left(\ \ |\epsilon+4n)\in {\cal H}^{\rm kin}\
:\ n\in \mathbb{Z}\ \right),\label{hnu}
\end{equation}
where $\epsilon$ is an arbitrary real number. We have the following
orthogonal decomposition into preserved subspaces:
\begin{equation}
{\cal H}^{\rm kin}\ =\ \overline{\bigoplus_{\epsilon}{\cal H}_{\epsilon}}.
\end{equation}
\bigskip

\noindent{\bf Discreteness.} Since  physicists often mean some
weaker definitions of the discreteness, let us be very precise here:
given an essentially self-adjoint operator $X$ defined in some
domain ${\cal D}$ in the Hilbert space ${\cal H}$, we say its
spectrum is discrete whenever the following conditions are
satisfied:
\begin{itemize}
\item there exists a basis of ${\cal H}$ consisting of the eigenvectors of $X$,
\item for each eigenvalue the corresponding eigenvectors  span a
{\it finite} dimensional subspace,
\item for every finite interval $I$ of $\R$, the set of the eigenvalues
of $X$ contained in $I$ is finite.
\end{itemize}

Going back to the case at hand:
\medskip

\noindent{\bf Proposition 3 } {\it For each of the subspaces ${\cal
H}_{\epsilon}$,  the restricted operator $\hat{C}_{\rm sc}^{\rm grav}:
{\cal H}_{\epsilon}\rightarrow {\cal H}_{\epsilon}$ considered as an
essentially self-adjoint operator in the Hilbert space
$\overline{{\cal H}_{\epsilon}}$ has a discrete spectrum.}
\bigskip

\noindent{\bf Estimate on the number of  eigenvectors} We need some
more notation. Given a self-adjoint operator $X$ in a Hilbert space
${\cal H}$, a number $\lambda\in\R$ and inequality relation
$$\iota\ =\ >,\,<,\,\le,\,\ge$$
we will denote by
$${\cal P}_{X\iota\lambda}:{\cal H}\rightarrow{\cal H}$$
the spectral projector of $X$ onto the interval
$\{x\in\R\ :\ x\iota\lambda\}$.
 The image will be denoted as follows
$$ {\cal H}_{X\iota\lambda}\ :=\ {\cal P}_{X\iota\lambda}({\cal H}).
$$

We are in the position to state our next result:
\medskip

\noindent{\bf Proposition 4 } {\it For each of the subspaces ${\cal
H}_{\epsilon}$,  the restriction operator $\hat{C}_{\rm sc}^{\rm grav}:
{\cal H}_{\epsilon}\rightarrow {\cal H}_{\epsilon}$ considered as an
essentially self-adjoint operator in the Hilbert space
$\overline{{\cal H}_{\epsilon}}$ satisfies:} $$ {\rm dim} {\cal
H}_{C_{\rm sc}>-E}\ \  \le\ \ {\rm dim} {\cal H}_{\tilde{A}\ge-E}\ \
\le\ \  {\rm dim} {\cal H}_{A' \ge-E}$$ {\it for arbitrary} $E>0$
{\it where}
\begin{align}
A'\ &:=\ - \frac{3\ell_0^2}{32 \pi G}\widehat{\sqrt{|p|}}\ :\ {\cal
H}_{\epsilon}\ \rightarrow\
{\cal H}_{\epsilon}\\
\tilde{A}\ &:=\
 -
\frac{3\ell_0^2}{8 \pi G \gamma^2}\left(-
\widehat{\sqrt{|p|}}\widehat{\frac{\sin^2\,(\frac{1}{2}\bar{\mu})}
{\bar{\mu}^2}}+
\frac{1}{4}(1+\gamma^{2})\widehat{\sqrt{|p|}}\right)\ :\ {\cal
H}_{\epsilon}\ \rightarrow\ {\cal H}_{\epsilon},
\end{align}
\medskip

Since the operators $A'$ and $\tilde{A}$ act just by multiplication by
functions (see (\ref{sqrtp}), the numbers ${\rm dim} {\cal
H}_{\tilde{A}>-E}\ \le {\rm dim} {\cal H}_{A' >-E}$ can be
calculated in a straightforward way, for each $E>0$.
\bigskip

\noindent{\bf Proof of Proposition 1} To show the essentially
self-adjointness of the operator $\hat{C}_{\rm sc}^{\rm grav}$ on
the domain ${\cal D}$ , we use (\ref{C2}) to present the operator in
the form
\begin{equation}
\hat{C}_{\rm sc}^{\rm grav} \ =\ -\frac{3\ell_0^2}{32\pi G\gamma^2}(
C_0 + H_1 ),
\end{equation}
where the function operator  $C_0$ is the same as in (\ref{C2}).
Obviously, the operator $C_0$ is essentially self-adjoint in the
domain ${\cal D}$. Therefore, it would be enough to show that
\begin{equation}\label{kato}
\|H_1\Psi\|^2\ \le\ \|C_0\Psi\|^2\ +\ \beta\|\Psi\|^2,
\end{equation}
for some constant $\beta$ and every $\Psi \in {\cal D}$, to conclude
that also the operator $C_0+H_1$ is essentially self-adjoint
(\cite{20} V.4.6). Applying the
inequality
$$ \|v+w\|^2\ \le\ 2\|v\|^2 + 2\|w\|^2,$$ true for arbitrary pair
of vectors, elements of a Hilbert space, one can check that
\begin{equation} \|H_1\Psi\|^2\ = \|(U_4C_4 \ +\
U_{-4}C_{-4})\Psi\|^2\ \le\ \langle\ \Psi\ |\
2(|C_4|^2+|C_{-4}|^2)\Psi\ \rangle,
\end{equation}
where the function $|C_4|^2+|C_{-4}|^2$ acts by multiplication as an
operator in ${\cal D}$. However, it follows from (\ref{sqrtp},
\ref{mubar}), that the function $(A(\nu)/\bar{\mu}^2(\mu(\nu)))$ is
linear in $\nu$ for $\nu\le -1$ as well as for $\nu \ge 1$.
Therefore it is easy to see, that
$$ C_0^2(\nu)\ =\ 2(|C_4|^2+|C_{-4}|^2)(\nu) + f_0+ f_1(\nu) +
f_2(\nu)$$ where $f_0$ is a constant, $f_1$ is some function of a
compact support, and $f_2(\nu)\ge 0$ for every $\nu$. That form is
sufficient to conclude the condition (\ref{kato}).
\bigskip

\noindent{\bf Proof of Proposition 2}
%
Let us remind that $\hat{C}_{\rm sc}^{\rm grav}$ is essentially
self-adjoint on the domain ${\cal D}$. For every $\Psi_0\in {\cal
D}(\hat{C}_{\rm sc}^{\rm grav})$, there exist sequence $\Psi_n\in
{\cal D}$, \be\Psi_n\rightarrow\Psi_0,\ \ \ \hat{C}_{\rm sc}^{\rm
grav} \Psi_n\rightarrow \hat{C}_{\rm sc}^{\rm grav}
\Psi_0.\label{phin}\ee \emph{Remark:  This kind of convergence is
referred as convergence in the graph norm
$$\|\Psi\|_B^2=\|B\Psi\|^2+\|\Psi\|^2,\ \ \ \Psi\in {\cal D}(B),$$
for operator $B$.}
\medskip

Suppose $\hat{C}_{\rm sc}^{\rm grav} \Psi_0=0$. We will prove
that the only possibility could be $$ \Psi_0\ =\ b|0),\ \ \ \ \ \
\ \ \ \ b\in\C. $$ But that possibility is ruled out by checking
by inspection, that $$ (0|\hat{C}_{\rm sc}^{\rm grav} |0)\ <\ 0.
$$

Let $\Psi_n\in {\cal D}$, $n=1,...,\infty$ be a sequence such that
(\ref{phin}). We have:
 $$\langle \Psi_n |\hat{C}_{\rm sc}^{\rm grav} \Psi_n\rangle\ \le\
 -\frac{3\ell_0^2}{32\pi G}\langle
 \Psi_n|\widehat{\sqrt{|p|}}\Psi_n\rangle \le
0$$ and we know that $$\lim_{n\rightarrow \infty}\langle \Psi_n
|\hat{C}_{\rm sc}^{\rm grav} \Psi_n\rangle\ =  0.$$ The comparison
implies
 $$
0\ = \  \lim_{n\rightarrow \infty}\langle
\Psi_n|\widehat{\sqrt{|p|}}\Psi_n\rangle\ =\ \lim_{n\rightarrow
\infty}\big\|\sqrt{\widehat{\sqrt{|p|}}}\Psi_n\big\|^2.$$
The second equality,  due to the closedness of the domain of
$\sqrt{\widehat{\sqrt{|p|}}}$, implies that the vector $\Psi_0$
belongs to the domain of the self-adjoint extension, $\Psi_0 \in
{\cal D}\big(\sqrt{\widehat{\sqrt{|p|}}}\big)$. In the consequence,
$$\sqrt{\widehat{\sqrt{|p|}}}\Psi_0=0.$$ The only solution is
$$\Psi_0=b|0).$$ That completes the proof.
\bigskip

\noindent{\bf Proofs of Proposition 3 and 4} The
propositions are a consequence of a single lemma, we formulate and
prove below (compare \cite{JD} XIII\footnote{We thank Jan Derezi\'nski for an important suggestion.}).
\medskip

\noindent{\bf Lemma } {\it Let ($A$,${\cal D}(A)$) and ($B$,${\cal
D}(B)$) be operators in a Hilbert space ${\cal H}$ with their
domains and ${\cal D}\subset {\cal D}(A)\cap {\cal D}(B)$ be a dense
subspace of ${\cal H}$.
Suppose the following conditions are satisfied:
\begin{itemize}
\item On the domain ${\cal D}$ the following inequality holds
$$ 0\le A\le B,$$
\item The operator $B$ is essentially self-adjoint in ${\cal D}$,
\item $A$, as an operator defined in ${\cal D}(A)$, is self adjoint,
positive and has discrete spectrum.
\end{itemize}
Then $B$ is also positive and has discrete spectrum. Moreover, the
following inequality holds for arbitrary $\lambda\ge 0$}
\[
{\rm dim}{\cal H}_{B<\lambda}\ \le\ {\rm dim}{\cal
H}_{A\le\lambda}. \label{est}
\]
\medskip

\begin{proof}
 Fix arbitrary $\lambda \ge 0$ and  the corresponding
 Hilbert space ${\cal H}_{B<\lambda}$. Given any  $\delta>0$ consider the
projection
\begin{equation}\label{proj}
P_{A\le\lambda+\delta}: {\cal H}_{B<\lambda}\rightarrow {\cal
H}_{A\le\lambda+\delta}.
\end{equation}
We  will show that its kernel is trivial. Then it is clear that
$${\rm dim}{\cal H}_{B<\lambda}
 \le {\rm dim}{\cal H}_{A\le\lambda+\delta}.$$
%
%
 Suppose $\Psi_0$ belongs to the kernel
of (\ref{proj}) and $\|\Psi_0\|=1$.
All the subspace ${\cal H}_{B\le\lambda}$ is contained in the domain
of the self-adjoint extension of $B$, that acts on this subspaces as
a bounded operator $BP_{0\le B\le \lambda-\delta'}$ with the upper
bound $\lambda$. Hence,
$$ \lambda \ge \langle \Psi_0 | B\Psi_0\rangle. $$
$\Psi_0$ may not be in the common domain ${\cal D}$. However, since
$B$ is essentially self-adjoint on ${\cal D}$, there is a sequence
of vectors $\Psi_n$,
$$\Psi_n\rightarrow\Psi_0,\ \ \ B\Psi_n\rightarrow B\Psi_0\ \ \ \
\ \Psi_n\in {\cal D}.$$
It follows that $$\langle \Psi_n | B\Psi_n\rangle\rightarrow \langle
\Psi_0 | B\Psi_0\rangle, \ \ \ \ \ {\rm as}\ \ \ n\rightarrow \infty
$$ On the other hand, for every $n$, we have
$$ \langle \Psi_n | B\Psi_n\rangle\ \ge \langle \Psi_n |
A\Psi_n\rangle.$$
On the right hand side of the inequality decompose
$$\Psi_n\ =\ \Psi_n^{\le}\ +\ \Psi_n^{>},\ \ \ \ \
\Psi_n^{\le}\in{\cal H}_{A\le\lambda+\delta},\ \ \Psi_n^{>}\in{\cal
H}_{A>\lambda+\delta}.$$
Note, that 
each $\Psi_n^{\le}$ belongs to the  domain of the self-adjoint
extension of  $A$, therefore so does $\Psi_n^>$.  Now,
\begin{equation}
\langle \Psi_n | B\Psi_n\rangle\ \ge\ \langle \Psi_n^{\le} |
A\Psi_n^{\le} \rangle\ +\ \langle \Psi_n^> | A\Psi_n^>\rangle\ \ge\
(\lambda+\delta)\|\Psi^>_n\|^2\label{key}
\end{equation}
But $\|\Psi^>_n\|^2 \ \rightarrow 1$ as $n\rightarrow\infty$, since
$\Psi_0=\Psi_0^>$, $\|\Psi_0\|=1$
and projection $P_{A>\lambda+\delta}$
is continuous. Finally, taking the limit of (\ref{key}) we find
$$\lambda\ge \lim_{n\rightarrow\infty}\langle \Psi_n | B\Psi_n\rangle\ge
\lambda+\delta.$$ The contradiction shows the kernel of projection
(\ref{proj}) is empty. However, since the spectrum of $A$ is
discrete, we have
$${\cal H}_{A\le\lambda+\delta}\ =\
{\cal H}_{A\le\lambda}$$ for sufficiently small $\delta>0$.
 That completes the proof of the inequality
(\ref{est}).
\end{proof}

\medskip

Proposition 2 follows from Lemma by fixing arbitrary $\epsilon\in \R$
and defining ${\cal H}$ to be the corresponding Hilbert space
$\overline{{\cal H}_{\epsilon}}$ and the subspace ${\cal D}\ :=\ {\cal
H}_{\epsilon}$, the domain of  the following operators
\begin{align} A\ =\ \widehat{\sqrt{|p|}}\ &:\ {\cal H}_{\epsilon}\
\rightarrow\
{\cal H}_{\epsilon}\\
B\ =\ -\frac{32 \pi\ G}{3\ell_0^2}\hat{C}_{\rm sc}^{\rm grav} \ &:\
{\cal H}_{\epsilon} \ \rightarrow\ {\cal H}_{\epsilon}.
\end{align}

A similar choice implies Proposition 4.

\section{Closing remarks and local difference between the
SU(2) and SO(3) universes.} Perhaps we should explain what  exactly
the initial technical difficulty  in the LQC closed FRW model was,
that we solved in this paper. In the previous attempts, the loop
used to replace the curvature component was somewhat complicated,
that made calculations unmanageable. We have found a neat analog of
a flat square in SU(2), that fits the framework much better (see the
contour $S_{ij}^\mu$ in Section 3.2).

We fixed a background metric tensor in SU(2) (SO(3)) corresponding
to the sphere of radius $2$. Unlike in the flat case, that metric
can be naturally distinguished. It provides the fiducial volume
$V_0$ and the parameter $\ell_0={V_0}^{1/3}$ (\ref{V_0}) present in
our formulae. The value of $\ell_0$ labels the SU(2) and
 the SO(3) cases. Note, that the "quantum of volume" $4V_1$ (\ref{nu1})
  defined by the quantum hamiltonian
 evolution is $\ell_0$ invariant. So one could think that the two
 cases are locally not distinguishable by the quantum evolution.
That  conclusion would not be correct. In fact, a given eigenvalue
of the volume operator (or, equivalently, the operator $\hat{p}$
(\ref{catmu}) defines on SU(2) the geometry locally different then
the geometry it defines on SO(3). Therefore the jump $4V_1$ in the
total volume has different meanings from the point of view of the
local geometry on SU(2) or SO(3) respectively. That shows, that even
making local observations of the geometry of our universe we should
be able to tell between the two cases.

The fact that $0$ is not an element of the spectrum of the scalar
constraint operator  of the gravitational field  for any of the
super selection sectors, means that there are no quantum vacuum
solutions (without the cosmological constant). That is in the
agreement with the classical theory.

Our results on the properties of the gravitational field scalar
constraint operator give a good insight  for the analysis of the
solutions of the full quantum scalar constraint of the gravitational
field coupled with the matter, with or without the cosmological
constant. We will demonstrate it in the coming paper.


\bigskip
\noindent{\bf Acknowledgements} {We thank Abhay Ashtekar for
explaining to us his improvement to the $k=0$ LQC model and drawing
our attention to the $k=1$ case.  We also benefited a lot from
discussions with Tomasz Pawlowski, Param Singh and Kevin
Vandersloot. Finally, Jan Derezinski gave us important suggestion
used in Lemma. The work was partially supported by the Polish
Ministerstwo Nauki i Szkolnictwa Wyzszego grant  1 P03B 075 29}

\end{document}